# COLLECTIVE INERTIA AND FISSION BARRIERS WITHIN THE SKYRME-HARTREE-FOCK THEORY

A. BARAN[a,b,c], A. STASZCZAK[a], J. DOBACZEWSKI[b,c,d], W.NAZAREWICZ[b,c,d]

[a] *Department Of Theoretical Physics, Institute of Physics, UMCS, ul. Radziszewskiego 10, 20-031 Lublin, Poland*

[b] *Department of Physics, University of Tennessee, Knoxville, TN 37996, Knoxville, USA*

[c] *Physics Division, Oak Ridge National Laboratory, P.O.Box 2008, Oak Ridge, TN 37831, USA*

[d] *Institute of Theoretical Physics, Warsaw University, ul. Hoża 69, Warsaw, Poland*

Spontaneous fission barriers, quadrupole inertia tensor, and zero-point quadrupole correlation energy are calculated for $^{252,256,258}$Fm in the framework of the self-consistent Skyrme-Hartree-Fock+BCS theory. Two ways of computing collective inertia are employed: the Gaussian Overlap Approximation to the Generator Coordinate Method and cranking *ansatz*. The Skyrme results are compared with those of the Gogny-Hartree-Fock-Bogolyubov model.

## 1. Introduction

The microscopic description of the fission process is one of the most challenging issues in nuclear structure. It represents an extreme example of the tunneling of the many-body system. There exist many descriptions of spontaneous fission based on the adiabatic assumption. While the microscopic-macroscopic models still offer the best fit to the data, during recent years, a number of self-consistent approaches rooted in the Adiabatic Time-Dependent Hartree-Fock method (such as the collective Schrödinger equation with microscopic mass tensor) have been introduced.

The spontaneous-fission half-life can be reliably estimated within the semi-classical WKB approximation. Here, the key quantity is the action integral

$$S = \int_{(s)} \left\{ 2\left[V(q) - E\right] \sum_{ij} B_{ij}(q) \frac{dq_i}{ds}\frac{dq_j}{ds} \right\}^{1/2} ds \tag{1}$$

along the trajectory $s$ in the multidimensional space of collective parameters $\{q\}$. In Eq. (1), $V(q)$ represents the collective potential given by the HF+BCS or HFB energy, $E$ is the ground-state energy and $B(q)$ is a second rank tensor of a collective mass.

The total energy of the system as a function of collective coordinates represents a zero-order approximation to the potential energy surface for shape-vibrations, $V(q)$. However, before one can use $V(q)$ in calculations with the collective Hamiltonian, dynamic corrections have to be added. The reason is that the underlying states have a finite uncertainty in the collective deformation. As a consequence, the potential $V(q)$ contains contributions from quantum fluctuations, and these contributions need to be subtracted first before adding the energies associated with the true physical zero-point fluctuations, $E_0$, see Ref.[1] The theoretical evaluation of these correction terms can be done in the framework of the cranking approximation (CRA), the generator coordinate method (GCM),[2] or the Gaussian overlap approximation to GCM (GOA).[3,4,5,6,7]

Having minimized the action $S$ in the collective space $\{q\}$, the spontaneous-fission half-life $T_{\rm sf}$ can be calculated from[8]

$$\log T_{\rm sf} = -20.54 + \log\left[1 + \exp\left(2S_{\rm min}\right)\right] - \log\left(2E_0\right). \tag{2}$$

It immediately follows from Eq. (1) that both the potential $V(q)$ and the inertia tensor $B(q)$ determine the fission dynamics in the multidimensional collective space; hence the value of $T_{\rm sf}$.

In this work we discuss fission barriers, mass parameters, and zero-point correlation energy (ZPCE) calculated in the Hartree-Fock+BCS model with the commonly used Skyrme SkM$^*$ and SLy4 energy density functionals. The Skyrme results are compared to those within the Gogny D1S Hartree-Fock-Bogolyubov model.[9,7] Another comparison is done between collective inertia obtained in the GOA and CRA approximations.

## 2. The Model

In order to make a comparison between various methods and models transparent, in the present calculations we only consider one collective variable $q$, an average axial quadrupole moment of the mass distribution in the nucleus:

$$q = Q_{20} \equiv \langle\phi|\hat{Q}_{20}|\phi\rangle, \tag{3}$$

where

$$\hat{Q}_{20} = \sum_{i=1}^{A} \sqrt{\frac{16\pi}{5}} r_i^2 Y_{20}(\theta_i, \phi_i) = \sum_{i=1}^{A} (2z_i^2 - x_i^2 - y_i^2)\,. \tag{4}$$

To account for pairing correlations, we applied the BCS approximation to the density-dependent $\delta$-pairing interaction (DDDI):

$$v_{\rm pair}(\vec{r}_1 - \vec{r}_2) = v_{0t}\delta(\vec{r}_1 - \vec{r}_2)\left[1 - \frac{\rho(\vec{r}_1)}{\rho_0}\right], \tag{5}$$

where $t$=p/n for protons/neutrons, $\rho = \rho(\vec{r})$ is the isoscalar nucleonic density, and $\rho_0$=0.16 fm$^{-3}$. The coupling constants $v_{0n}$=842 MeV Fm$^3$ and $v_{0p}$=1020 MeV fm$^3$ were fitted to the experimental pairing gaps in $^{252}$Fm: $\Delta_n = 0.696$ and $\Delta_p = 0.803$ MeV. The pairing-active space consisted of the lowest $Z$ ($N$) proton (neutron) single-particle states. We have also performed calculations with the seniority pairing force (with the constant strength parameters) given by

$$\begin{aligned} G_n &= \left[24.70 - 0.108\,(N - Z)\right]/A\,, \\ G_p &= \left[14.76 + 0.241\,(N - Z)\right]/A. \end{aligned} \tag{6}$$

The calculations were performed using the code HFODD (v2.19l)[10,11,12] that allows for an arbitrary symmetry breaking. For the basis, we took the lowest 1140 single-particle states of the deformed harmonic oscillator. This corresponds to 14 oscillator shells at the spherical point. Based on the HFODD self-consistent wave-functions, the collective mass tensor components and ZPCE were computed. In principle, all the collective-tensor components $\lambda\mu$ with multipolarities $\lambda = 2, 3, \ldots, 9$ can be calculated.

In this study, only specific static fission paths are considered; they have been obtained in the calculations presented in Ref.[13] Specifically, for the fermium isotopes considered, one predicts both reflection-symmetric (s) and reflection-asymmetric (a) fission valleys. There are two kinds of symmetric paths predicted; namely, the valley that corresponds to elongated fission fragments (E), and that with more compact (C) fragments that resemble spherical $^{132}$Sn clusters when approaching $^{264}$Fm. (The shorthand notation sEF, used in the following, means symmetric, elongated-fragments fission path; similarly aEF and sCF.)

## 3. Inertia Tensor and Zero-Point Correlation Energy

In order to calculate the inertia tensor one usually applies the GOA or uses the cranking model formalism. In both cases, the main task is the calculation of $\boldsymbol{M}^{(k)}$,

$$M_{ij}^{(k)}(\boldsymbol{q}) = \sum_{\nu\nu'} \frac{\langle\phi|\hat{Q}_i|\nu\nu'\rangle\langle\nu\nu'|\hat{Q}_j^{\dagger}|\phi\rangle}{(E_\nu + E_{\nu'})^k}\,, \tag{7}$$

where $q_i = \langle\hat{Q}_i\rangle$ are the collective coordinates ($i$ stands for $\lambda\mu$), $|\nu\nu'\rangle = \alpha_\nu^\dagger\alpha_{\nu'}^\dagger|\phi\rangle$ is a two-quasiparticle state, and $E_\nu$ is the quasiparticle energy.

Both in the GOA and cranking, the inertia tensor can be given by compact expressions[3,4]

$$\boldsymbol{B}^{\mathrm{GOA}} = \boldsymbol{\Sigma}^{(2)}[\boldsymbol{\Sigma}^{(1)}]^{-1}\boldsymbol{\Sigma}^{(2)}\,, \tag{8}$$

$$\boldsymbol{B}^{\mathrm{CRA}} = \boldsymbol{\Sigma}^{(3)}\,, \tag{9}$$

where matrices $\boldsymbol{\Sigma}^{(k)}$ read

$$\boldsymbol{\Sigma}^{(k)} = \frac{1}{4}\boldsymbol{M}^{(1)-1}\boldsymbol{M}^{(k)}\boldsymbol{M}^{(1)-1}\,. \tag{10}$$

In the nuclear case two kinds of fermions are present. The total covariant inverse inertia for a composite system is given as a sum of proton and neutron covariant inertia tensors.[3] This leads to the final expression

$$B = \gamma^2 \frac{B_n B_p}{\gamma_p^2 B_n + \gamma_n^2 B_p}, \tag{11}$$

where $\gamma$ is the total metric tensor, which is a sum of proton ($\gamma_p$) and neutron ($\gamma_n$) contributions.

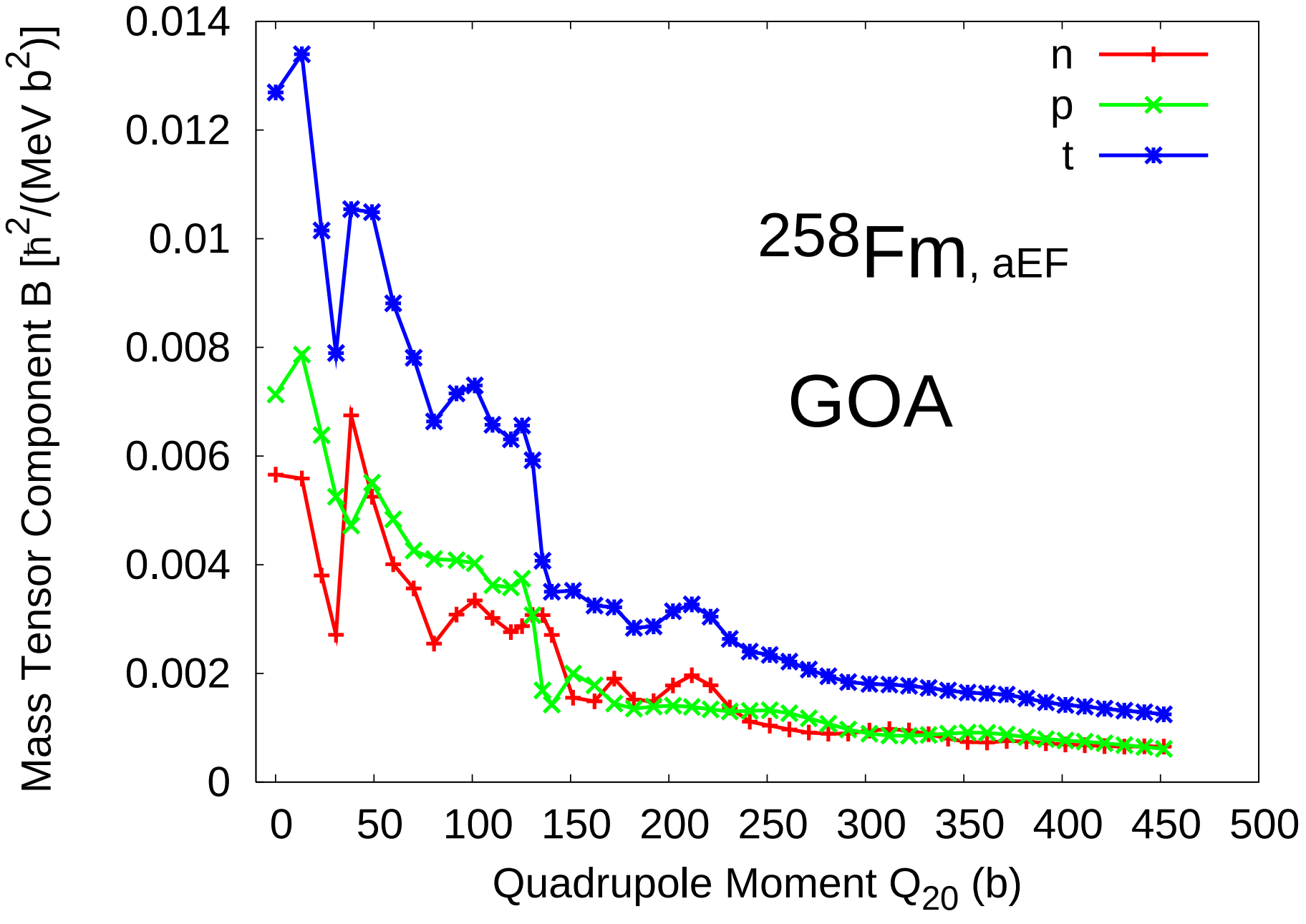

Fig. 1. The total (t) GOA quadrupole collective inertia and its proton (p) and neutron (n) components calculated for $^{258}$Fm in the SkM* model with seniority pairing along the asymmetric elongated static path to fission.

We conclude this section by recalling that the zero-point correlation energy in the GOA can also be expressed through quantities (10),[4] i.e.,

$$E_0 = \frac{1}{4}\mathrm{Tr}(\mathbf{\Sigma}^{(2)-1}\mathbf{\Sigma}^{(1)}). \quad (12)$$

## 4. Results

We begin our discussion from Fig. 1, which shows the quadrupole GOA inertia $B_{20}$ (8) for $^{258}$Fm calculated along the aEF path in the SkM*+seniority model. The corresponding proton and neutron contributions are also depicted; they are related to the total inertia through Eqs. (10) and (11). The behavior of the total inertia can easily be traced back to the proton and neutron shell effects that are responsible for characteristic fluctuations in the collective mass.

Figure 2 compares the values obtained in SLy4+DDDI and D1S models for $^{252}$Fm along the sCF static fission path. While the SLy4 values usually exceed those obtained with the Gogny force by more than a factor of two, the general patterns of $B_{20}(Q_{20})$ are fairly similar. The same conclusion holds for comparison between GOA and cranking inertia, with the cranking value being larger. A very interesting feature of the self-consistent inertia parameters is their regular behavior at large elongations. This has not been seen in earlier calculations using phenomenological potentials (see, e.g., Ref.,[14]) where the oscillating behavior of $B_{20}$ persisted in the whole deformation range.

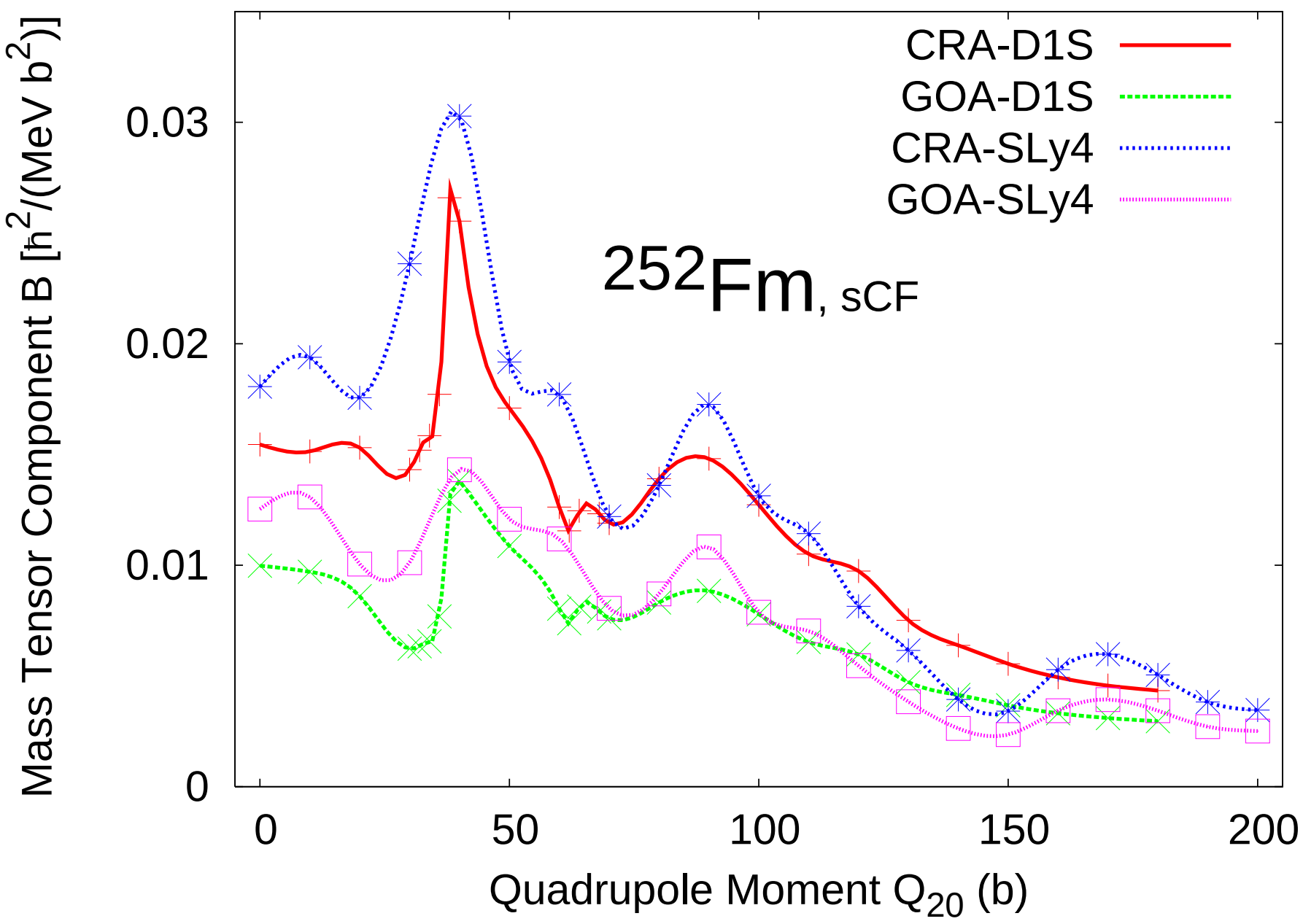

Fig. 2. Comparison of $B_{20}$ in GOA and CRA for $^{252}Fm$ calculated along the symmetric compact path to fission in Skyrme-HF (SLy4+DDDI) and Gogny-HF (D1S) models.

In order to calculate the collective potential $V(q)$ and fission barriers, the zero-point correlation energy has to be evaluated first. The ZPCE (12) for $^{252}$Fm calculated in SLy4+DDDI and D1S is shown in Fig. 3. One sees a qualitative and quantitative agreement between both variants of calculation. The largest correction is predicted around the first minimum ($Q_{20}$≈25 b). As seen in Fig. 4, the inner and outer fission barriers in $^{252}$Fm are significantly modified by the ZPCE. Namely, the barriers calculated with the corrected collective potential are ∼1MeV higher in the whole range of $Q_{20}$.

Finally, Figs. 5 and 6 display the collective potentials and inertia parameters (in GOA and CRA) in $^{256}$Fm and $^{258}$Fm, respectively, calculated along the symmetric (sEF and sCF) and asymmetric (aEF) fission paths.[13] These results confirm the previous observations; namely, the deformation patterns of collective inertia in CRA and GOA are very similar, with the cranking values being appreciably higher.

## 5. Summary

In this study we performed pilot calculations of collective inertia, zero-point quadrupole correlation energy, and fission barriers in selected Fermium isotopes.

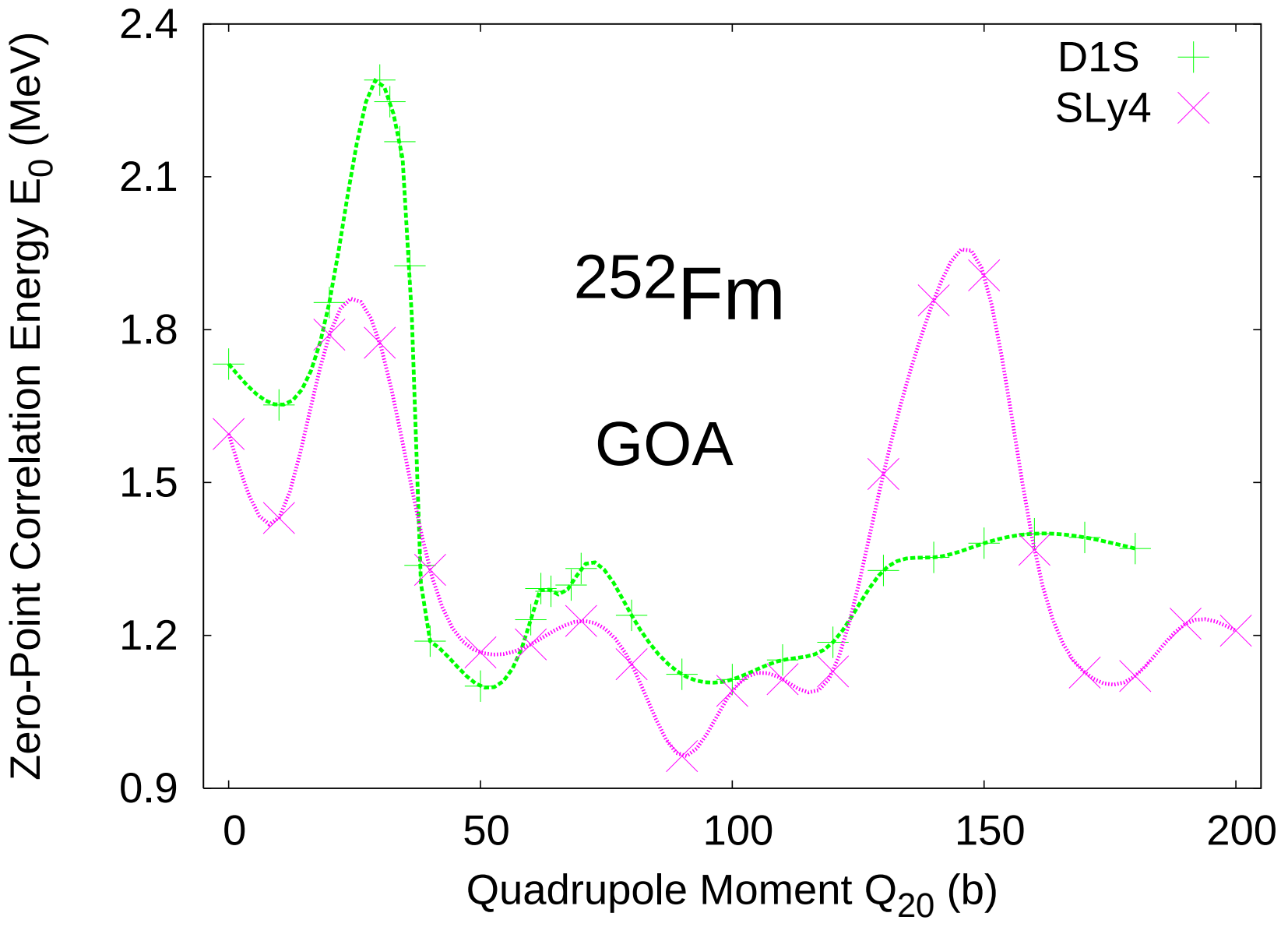

Fig. 3. The zero-point correlation energy (12) in GOA along the aEF fission path in $^{252}$Fm calculated in SLy4-DDDI (+) and D1S (X) models.

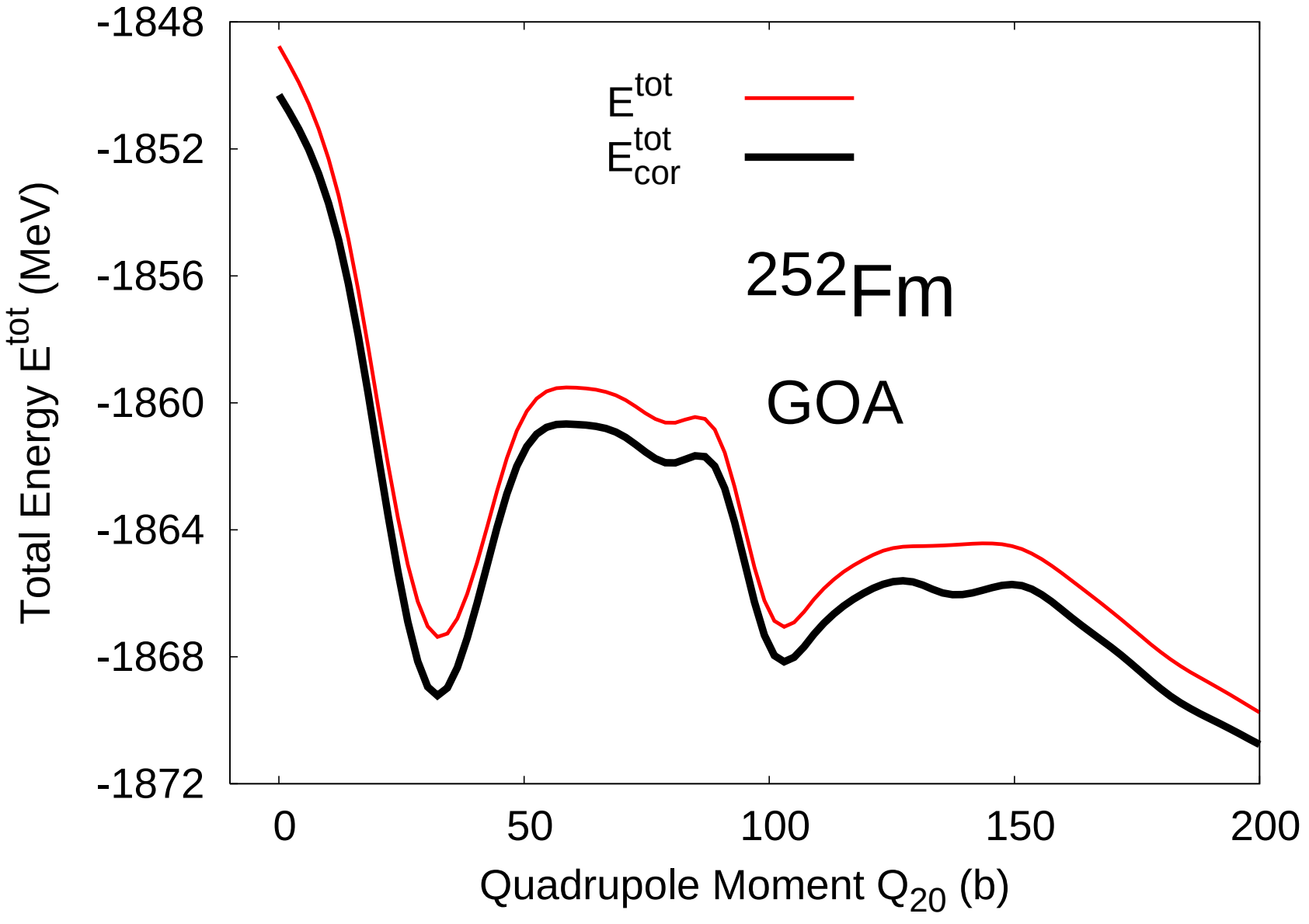

Fig. 4. The collective potential in $^{252}$Fm in SLy4-DDDI along the aEF fission path without ($E^{\rm tot}$; thin line) and with ( $E^{\rm tot}_{\rm cor} - E_0$; thick line) the ZPCE included.

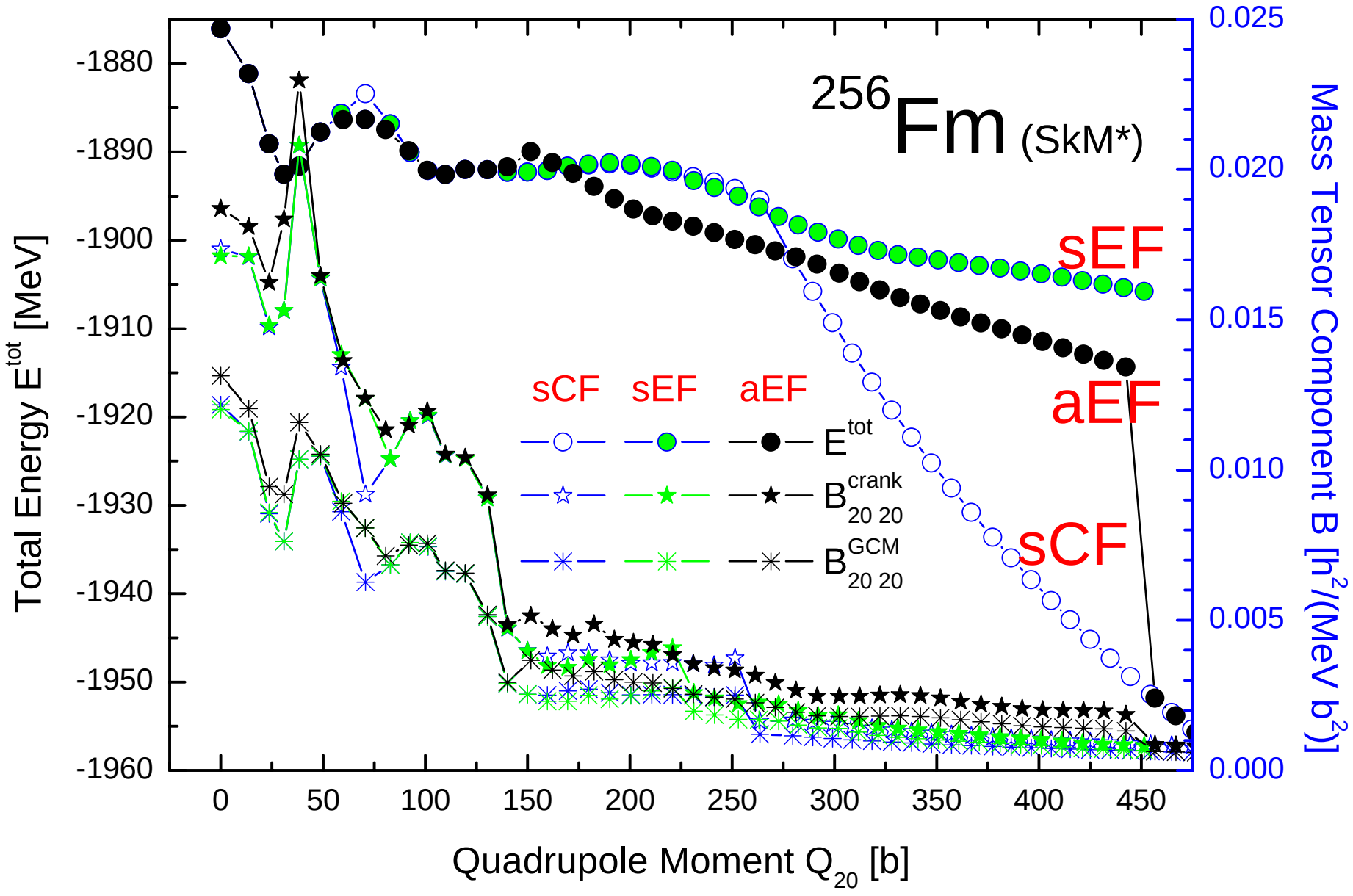

Fig. 5. The collective potential (left-hand-side scale) and corresponding GOA and CRA quadrupole inertia (right-hand-side scale) along three fission paths in $^{256}$Fm using the SkM*+seniority model.

The HFODD code employed allows for an arbitrary symmetry breaking; this feature is of crucial importance when discussing spontaneous fission where the reflection-asymmetric and triaxial shapes can play a role. The main conclusion of this work can be summarized as follows:

- The zero-point correlation energy is important to include as it can significantly modify the action by lowering the ground state by approximately 1 MeV in the nuclei considered.
- The collective quadrupole inertia calculated in the Skyrme SLy4 and SkM* models have very similar deformation pattern to those in the Gogny D1S model. Our HF-Skyrme values are very close to those of Refs.[7,9]
- The collective inertia obtained in the cranking approximation are about twice as large as the GOA results. In all the cases considered, the collective inertia smoothly decrease at large deformations; they do not exhibit oscillations seen in previous microscopic-macroscopic calculations.
- As discussed earlier, both $V(q)$ and $B(q)$ contribute to the action integral $S$ (1). In general, the fission barriers in the Gogny model are higher as compared to those in SkM* and the opposite holds for the collective inertia.

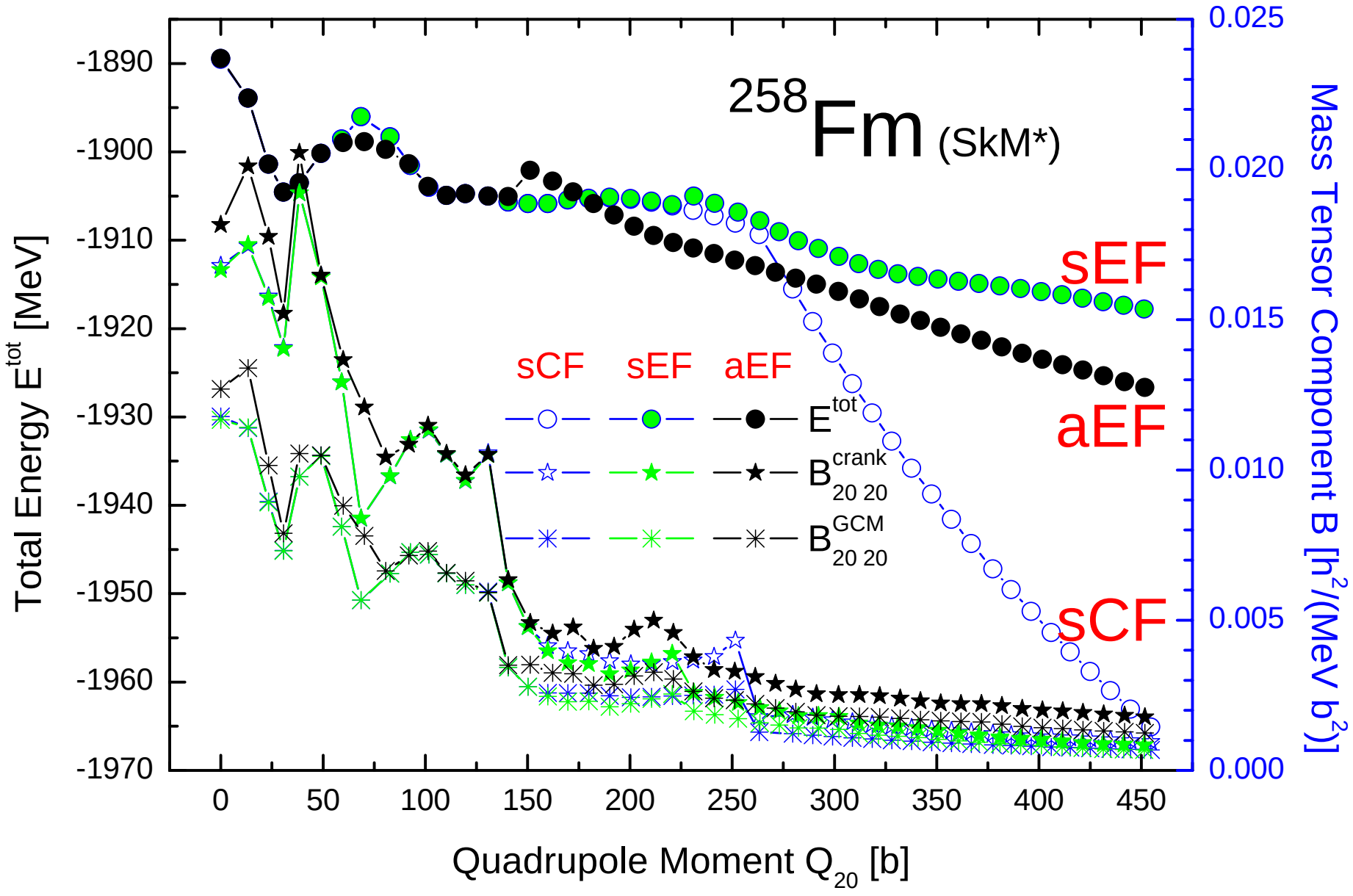

Fig. 6. Similar to Fig. 5 except for $^{258}$Fm.

However, both those effects partly cancel in the product $B(V - E)$ that determines the action and the spontaneous fission lifetime.

## 6. Acknowledgements

We are indebted to Michał Warda for making his Gogny-HF (D1S) results available to us and to Leszek Próchniak for valuable discussions. This work was supported in part by the National Nuclear Security Administration under the Stewardship Science Academic Alliances program through the U.S. Department of Energy Research Grant DE-FG03-03NA00083; by the U.S. Department of Energy under Contract Nos. DE-FG02-96ER40963 (University of Tennessee), DE-AC05-00OR22725 with UT-Battelle, LLC (Oak Ridge National Laboratory), and DE-FG05-87ER40361 (Joint Institute for Heavy Ion Research); by the Polish Committee for Scientific Research (KBN) under Contract No. 1 P03B 059 27; by the Foundation for Polish Science (FNP); and by the Polish Ministry of Science and Education under Contract N 202 179 31/3920.